\documentclass{JAC2003}

\usepackage{graphicx}
\def\be{\begin{eqnarray} &&} 
\def\nonu{\nonumber \\ &&} 
\def\ee{\end{eqnarray}} 
\def\psla{\slash \! \! \!} 

\begin{document}
\title{ELECTROMAGNETIC FORM FACTORS OF HADRONS \\ IN THE SPACE- AND TIME-LIKE
 REGIONS\\ WITHIN A CONSTITUENT
QUARK MODEL ON THE LIGHT FRONT }

\author{J. P. B. C. de Melo, Inst. de F\'\i sica Te\'orica, Univ. Est. 
Paulista, 
01405-900, S\~ao Paulo,  Brazil,\\
 Tobias Frederico, Dep. de F\'\i sica, ITA-CTA, 12.228-900, S\~ao Jos\'e dos 
Campos, Brazil, \\
Emanuele Pace, Dip. di Fisica, Univ. di Roma "Tor Vergata" and INFN
 Sez. Tor Vergata, \\Via della Ricerca 
Scientifica 1, I-00133  Roma, Italy, \\
Giovanni Salm\`e, INFN, Sez. Roma I, P.le A. Moro 2, 
 I-00185 Roma, Italy 
}

\maketitle

\begin{abstract} An approach for describing the electromagnetic 
form factors of hadrons,  in the space- and time-like
 regions, within a constituent
quark model on the light front is shortly illustrated
 and  calculations for 
the pion
case are reported. Three main ingredients enter our approach: i) 
the dressed photon vertex 
where a photon decays in a quark-antiquark pair, ii)  the  
on-shell quark-hadron vertex functions in the valence sector, and iii) the non-valence
component of the hadron state.
\end{abstract}

\section{INTRODUCTION}
 The   experimental study of hadron electromagnetic (em) form factors  allows 
 us  to gather valuable  information on  hadron states, and the increasing
 accuracy of the experiments calls for  refined theoretical tools in order
 to make stringent the interpretation of the data. In particular, the
 analysis  of em form factors both in the space- and 
time-like regions, within the light-front  framework \cite{brodsky}, opens a unique 
possibility to study
 hadron states in  the valence sector and in the nonvalence one. 
 As a matter of fact, one can write the states for mesons and
baryons, respectively, as follows

\be
| meson \rangle =  |q\bar{q} \rangle + |q \bar{q} q \bar{q}\rangle +
|q \bar{q} ~g\rangle +
 ..... 
\nonu
| baryon \rangle = |qqq \rangle + 
|qqq~q \bar{q} \rangle +|qqq~g \rangle +
 ..... \nonu 
\ee
Indeed, it should be pointed out that only within the light-front approach
  the Fock
expansion becomes meaningul, given the coincidence between the mathematical
vacuum and the physical one.

 As it is well-known, investigations of the em properties of hadrons
 in the time-like (TL) region  yields the
 possibility to
address a vast phenomenology, and consequently  to impose
 quantitative constraints 
 on dynamical models pointing to a  microscopical description of hadrons.
 
\section{GENERAL FORMALISM}
The Mandelstam formula for the matrix elements of the em current \cite{mandel} is the starting point
for the contruction of our approach. Following Ref. \cite{mandel},
 the matrix elements in the TL region
read
\be
\hspace{-0.1in}j^{\mu} = -\imath e\int
\frac{d^4k}{(2\pi)^4} Tr\left [S_Q(k-P_{\bar h}){\bar \lambda}_{ h}(k-P_{\bar h},P_{  h})
 \right.\nonu \left.   
 S(k-q)~\Gamma^\mu(k,q)~S(k)~{\lambda}_{\bar{h}}(k,P_{\bar h}) 
 \right]
\label{mand}   
\ee 
where
 $\displaystyle
S(p)=\frac{1}{\psla p-m+\imath \epsilon} \,$, 
with $m$ the mass of the constituent quark struck by the virtual photon,
 $ S_Q(p)$ is the propagator of
 the spectator constituent, quark or diquark (in a simple
picture of baryons),
 $\Gamma^\mu(k,q)$  the quark-photon vertex, $q^{\mu}$  the 
virtual photon momentum,  $\overline
\lambda_{h}(k,P_{h})$   the hadron vertex function, $P^{\mu}_{h}$ and
 $P^{\mu}_{h^{\prime}}$ are the hadron momenta. The "bar" notation on the vertex function means that the associated 
amplitude is the solution of the Bethe-Salpeter equation, where the 
irreducible kernel is placed on the right of the 
amplitude, while in the conventional case it is placed on the left 
of the Bethe-Salpeter amplitude \cite{lurie}.

  For the space-like (SL)
 region, the crossing symmetry has to be applied, replacing the antihadron
 vertex with the corresponding hadron one, reverting the sign of the
 four-momenta and appending the proper number of $\gamma^5$.

\section{THE PION EM FORM FACTOR}
For the sake of concreteness, let us consider the case of the pion. If
both the SL   and TL regions have to be investigated, it is
necessary to perform the analysis of the em form factor in a frame where the
plus component of the four-momentum transfer is nonvanishing,  $q^+\neq 0$ 
\cite{ba01,pach02}. Following \cite{LPS} one can choose $q^+\neq 0$
and ${\bf q}_{\perp}=0$ . 
Figs. 1 and 2 \cite{plbpion} show a diagrammatic analysis of the physical
processes pertaining to SL and TL regions, respectively. In particular, 
Figs. 1(b) and 2(b) illustrate the contribution due to  
a $q\bar{q}$ pair, produced by the virtual photon. Such a contribution is 
dominant for both kinematical regions, if a vanishing pion mass is assumed. 
In what follows, in order to emphasize the unified description of
the em form factor in TL and SL regions, the simplifying assumption of a chiral
pion ($m_\pi =0$) will be adopted. From figs. 1 and 2   the
 following questions immediately  arise : i) how to model the quark-photon 
vertex ? ii)
 how to deal with   the amplitude for the
emission or absorption of a pion by a quark ?  iii) how to describe  the
$q\bar{q}$-pion vertex? The illustration   of our 
proposal for the construction of a phenomenological 
model in order to
answer such questions is the aim of the present contribution.  
\parbox{3.35in}{\includegraphics*[width=3.35in]{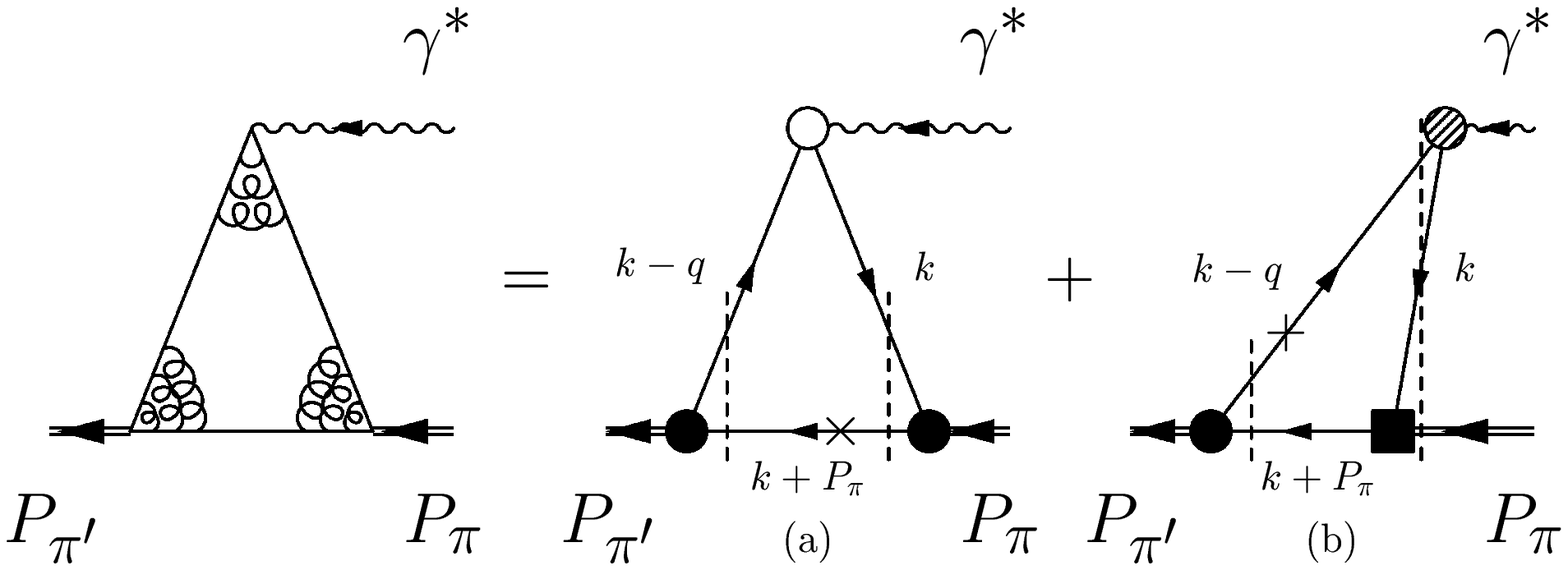}

\vspace{-2.9in}

Figure 1. Diagrammatic representation of the pion 
elastic form factor for  $q^+ >0$ vs the global $x^+$-time flow. 
 Diagram $(a)$ 
 is the contribution of the 
valence component in the initial pion wave function. 
Diagram $(b)$  is 
the non-valence contribution to the pion form factor. 
The crosses correspond to the  quarks on the $k^-$ shell. (After
\cite{plbpion}). \medskip \medskip
}

\hspace{-0.15in}\parbox{3.35in}{
\includegraphics*[width=3.3in]{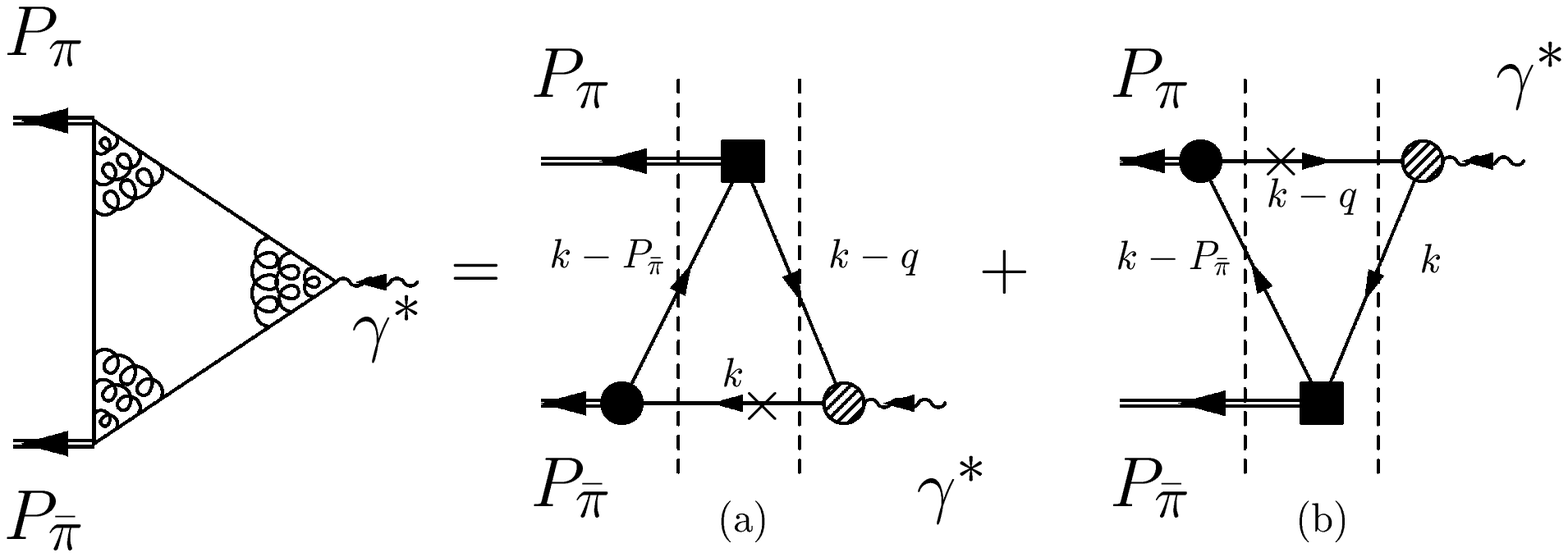}

\vspace{-2.9in}
Figure 2. Diagrammatic representation of the photon decay 
($\gamma ^* \rightarrow \pi \bar{\pi}$) 
 vs the global $x^+$-time flow. 
Diagrams (a) and (b) correspond to different  $x^+$-time orderings. 
The crosses correspond to the  quarks on the $k^-$ shell. (After
\cite{plbpion}).\medskip \medskip}
For the pion, Eq. (\ref{mand}) becomes
\be
\hspace{-.15in}j^{\mu} =-\imath 2 e \frac{m^2}{f^2_\pi} N_c\int
\frac{d^4k}{(2\pi)^4}
{\bar \Lambda}_{{\pi}}(k-P_{\bar  \pi},P_{{\pi}})
 \Lambda_{\bar \pi}(k,P_{\bar \pi})   
\nonu 
Tr[S(k-P_{\bar \pi}) \gamma^5
S(k-q)~\Gamma^\mu(k,q)~S(k) \gamma^5 ]
\label{jmu}   
\ee 
where  
$N_c=3$ is the number of colors and the factor $2$ comes from the isospin
weight.

Following Ref. \cite{plbpion}, a  Vector Meson Dominance approximation
 has been applied to the
quark-photon vertex, when a $q\bar{q}$ production is present. In
particular,  the
plus component of the quark-photon vertex  reads as follows \cite{tob}
  \be
\hspace{-.15in} \Gamma^+(k,q) =  \sum_{n, \lambda}~
\left [ \epsilon_{\lambda} \cdot \widehat{V}_{n}(k,k-q)  \right ]   
\Lambda_{n}(k,P_n) \nonu
{ \sqrt{2} [\epsilon ^{+}_{\lambda}]^* f_{Vn} \over (q^2 -
M^2_n + \imath M_n \Gamma_n(q^2))}
\label{vert}  
\ee
where $f_{Vn}$ is the decay constant of the n-th vector
meson (VM) into a virtual photon, 
 $M_n$($P_n$)  the mass (four-momentum) of the VM, 
 $\Gamma_n(q^2)=\Gamma_n q^2/M^2_n$ (for $q^2>0$) 
 the corresponding
 total decay width,
 $\epsilon_{\lambda}$  the VM polarization, and    
  $\left [ \epsilon_{\lambda} \cdot \widehat{V}_{n}(k,k-q)  \right ]  ~ 
\Lambda_{n}(k,q)$  the VM vertex function. In the actual calculation of 
$f_{Vn}$,
 the momentum component, $\Lambda_{n}(k,P_n)$, 
 of the VM vertex function, must be   
evaluated on the quark $k^-$ shell (i.e., for $k^- = k^-_{on} =
 (|{\bf k}_{\perp}|^2 + m^2)/k^+ $); in this case it   
  will be related for $0 < k^+ < P^+_n$ to the momentum part of the  
front-form VM wave function \cite{Jaus90}, which describes the 
 valence component of 
the meson state $|n\rangle$, viz.
\be 
\hspace{-.15in}\psi_{n}(k^+, {\bf k}_{\perp}; P^+_{n}, {\bf P}_{n \perp}) = ~ 
\left.
\Lambda_{n}(k,P_{n}) \right |_{[k^- = k^-_{on}]}~\times \nonu
\frac{P^+_{n} ~ ~ }{[M^2_n - M^2_0(k^+, {\bf k}_{\perp};
 P^+_{n}, {\bf P}_{n \perp})]} 
 \label{wfn}   
\ee
where $M_0$ is the standard front-form free mass and $\psi_{n}$ is the
eigenfunction, in the
  $1^-$ channel, of the square 
mass operator proposed in Refs. \cite{tobpauli,FPZ02}  within a relativistic  
constituent quark model. This model  takes into account the 
confinement through a harmonic oscillator potential and the  
$\pi-\rho$ splitting through a Dirac-delta interaction in the pseudoscalar  
channel. It achieves a  
satisfactory description of the experimental masses 
for both singlet and triplet $S$-wave mesons, with a  natural 
explanation of the "Iachello-Anisovitch law"~\cite{Iach,ani}, namely the almost-linear  
relation 
between the square mass  of the excited states and the radial quantum 
number $n$. Since the model of Refs. \cite{tobpauli,FPZ02}  
does not include the mixing between isoscalar and 
isovector mesons, in what follows  only  contributions from the isovector 
$\rho$-like vector mesons are included. 
 
The VM eigenfunction, $\psi_{n}(k^+, {\bf k}_{\perp}; q^+,{\bf
q}_{\perp})$,   
 which describes  the valence component of 
the VM state $|n\rangle$, is normalized to the probability 
of the lowest ($q\bar q$) Fock state (i.e. of the valence component). The $q\bar q$ 
probability can be roughly estimated to be $\sim 1/\sqrt{2n + 3/2}$ in a 
simple model \cite{tob} 
that reproduces the "Iachello-Anisovitch law" \cite{Iach,ani}, 
and is based on an expansion of the  
VM state $|n\rangle$, in terms of properly weighted 
Fock states $|i\rangle_0$, 
with $i > 0$ quark-antiquark pairs.

If in the calculation of the decay constant, $f_{Vn}$, 
the following assumptions on the analytic behaviour of the VM
 vertex function are adopted: i) $\Lambda_n(k,P_n)$
does not diverge in the complex plane $k^-$ for
$|k^-|\rightarrow\infty$, and ii)
  the contributions of its singularities 
 in the  integration over $k^-$ are negligible, one obtains
\be
\hspace{-.2in}f_{Vn} =  - {N_c \over 4 (2\pi)^3} \int_0^{P_n^+} \hspace{-.15in}
{dk^+ ~ d{\bf k}_{\perp}
\over k^+ ~ (P_n^+ - k^+)} \psi_{n}(k^+, {\bf k}_{\perp}; M_{n}, {\vec 0}_{\perp})
\nonu Tr \left [ (\psla k + m)  \gamma^+  (\psla k - \psla P_n + m) 
\widehat{V}_{n z}(k,k-P_n) 
 \right ]
 \label{fvn}  
\ee

The form factor of the pion in the 
TL and in the SL regions can be obtained from the plus 
component of the proper current matrix elements: 
$j^{\mu}_{TL} =\langle \pi \bar{\pi}| \bar{q} \gamma^{\mu}q 
|0\rangle = \left (P^{\mu}_{\pi} -P^{\mu}_{\bar{\pi}} \right )~F_{\pi}(q^2)$ , 
and
$j^{\mu}_{SL} =\langle \pi | \bar{q} \gamma^{\mu}q |\pi ^{\prime}\rangle =  
\left (P^{\mu}_{\pi} + P^{\mu}_{\pi ^{\prime}} \right )~F_{\pi}(q^2)$.  
Since in the limit $m_\pi \rightarrow 0$ the form factor 
 receives contributions only from the diagrams of Figs. 1(b) and 2(b),  
where the photon decays in a $q\overline q$ pair,  
one can apply our approximation  for the plus component 
of the dressed photon vertex (\ref{vert}), 
both in the SL and in the TL regions. 
Then the matrix element $j^+$ can be written as a sum  
 over the vector mesons and 
 consequently the form factor becomes 
 \be 
 \hspace{-.15in}F_{\pi}(q^2) = \sum_n~ {f_{Vn} \over q^2 - M^2_n + \imath M_n \Gamma_n(q^2)} ~ 
 g^+_{Vn}(q^2) 
\label{tlff}  
\ee 
where $g^+_{Vn}(q^2)$, for $q^2 > 0$, is the form factor for the VM decay in a pair of pions. 
 
Each VM contribution to the sum (\ref{tlff}) is 
 invariant under kinematical front-form boosts and 
therefore 
it can be evaluated in the rest frame of the  
corresponding resonance. 
 In this frame one has $q^+=M_n$ and $q^-=q^2/M_n$ for the photon and 
$P^{+}_n= P^{-}_n = M_n$  for the vector meson.   
  This means that we choose a different frame for 
each resonance (always with ${\bf q}_{\perp}=0$), 
 but all the frames are related by kinematical front-form boosts 
along the $z$ axis to each other, and to the frame where $q^+ = -q^- = \sqrt{-q^2}$ 
~ ($q_z=\sqrt{-q^2}$), adopted  in previous analyses of the SL region 
\cite{pach02,LPS}. 
Since in our reference frame one has 
$\sum_{\lambda} \left [ \epsilon 
^{+}_{\lambda}(P_n) \right ]^* \epsilon _{\lambda}(P_n) \cdot 
\widehat{\Gamma}_{n} = \left [ \epsilon ^{+}_{z}(P_n) \right ]^* 
\epsilon _{z}(P_n) \cdot \widehat{\Gamma}_{n} = - 
\widehat{\Gamma}_{n z}$, 
 we obtain \cite{plbpion} 
\be  
\hspace{-.15in}g^+_{Vn}(q^2) =  {N_c \over 8 \pi^3} { \sqrt{2} \over P^+_{\overline{\pi}} }  
\frac{m}{f_\pi} ~ \int_0^{q^+} {dk^+ \over (k^+)^2~(q^+-k^+)} \nonu\hspace{-.15in}\int d{\bf k}_{\perp} ~  
Tr \left [ \Theta^z ~ \left.
\Lambda_{\bar \pi}(k ; P_{\bar \pi})\right |_{(k^-=q^- + (k - q)^-_{on})} \right ] ~ 
\times 
\nonu \nonu \hspace{-.15in}   
\psi^*_{{\pi}}(k^+, {\bf k}_{\perp}; 
P^+_{{\pi}},{\bf P}_{{\pi} \perp })  
~\psi_{n}(k^+, {\bf k}_{\perp}; q^+,{\bf q}_{\perp})~
\times \nonu
\hspace{-.15in}{[M_n^2 - M^2_0(k^+, {\bf k}_{\perp}; q^+, {\bf q}_{\perp})]  
\over [q^2-M^2_0(k^+, {\bf k}_{\perp}; q^+, {\bf q}_{\perp})+i\epsilon]} 
\label{ffpi2} 
\ee  
where  $\Theta^z = {\mathcal V}_{nz}(k,k - q) 
~\gamma^5~ \left [\psla k - \psla P_{\bar \pi} + m \right ] ~ \gamma^5$.   
To obtain Eq. (\ref{ffpi2}) we have i)  
performed the $k^-$ integration assuming once more negligible contributions 
from the singularities in the vertex function, and ii) adopted Eqs. (\ref{wfn})
  for describing the quark-VM vertex in the valence sector.
The same assumption is
adopted for the momentum part of the $q\bar{q}$-pion vertex in the
 valence sector, namely
\be  
\hspace{-.15in}\psi_{\pi}(k^+, {\bf k}_{\perp}; P^+_{\pi}, {\bf 
P}_{\pi \perp}) = \frac{m}{f_\pi}
 \left. \Lambda_{\pi}(k,P_{\pi}) \right|_{[k^- = k^-_{on}]}~ 
\times \nonu\frac{P^+_{\pi}~  }
{[m^2_\pi - M^2_0(k^+, {\bf k}_{\perp}; P^+_{\pi}, {\bf P}_{\pi \perp})]}   
\label{wf2} 
\ee   
 It should be pointed out that we do not distinguish between
 $\left. \Lambda_{\pi(n)}(k,P_{\pi(n)}) \right|_{[k^- = k^-_{on}]} $ and 
$\left. \Lambda_{\pi(n)}(k,P_{\pi(n)}) \right|_{[k^- =P^-_{\pi(n)} - (P_{\pi(n)}-k)^-_{on}]}$, 
in the range $0 < k^+ <P^+_{\pi(n)}$, given the symmetry between the quark
momenta. Finally,  we have introduced a third main approximation:  
 following Ref. \cite{JI01}, 
 the momentum part of the quark-pion emission  vertex in the non-valence 
sector, 
$  \frac{m}{f_\pi} 
[{\Lambda} _{\bar \pi}(k ; P_{\bar \pi})]_{(k^-=q^- + (k - q)^-_{on})} $  
(see the square blob in Fig.  2(b)), is assumed to be a constant. 
 
The value of the constant is fixed by the pion charge normalization. 
 The same constant value is assumed for  
 the quark-pion absorption vertex (see the square blob in Fig. 1(b)). 
 
  It turns out that the same expression for $g^+_{Vn}(q^2)$ holds
 both in the TL and in the SL  
 regions \cite{plbpion}.
  
\subsection{Results}
\begin{figure}[t] 
\includegraphics[width=3.2in]{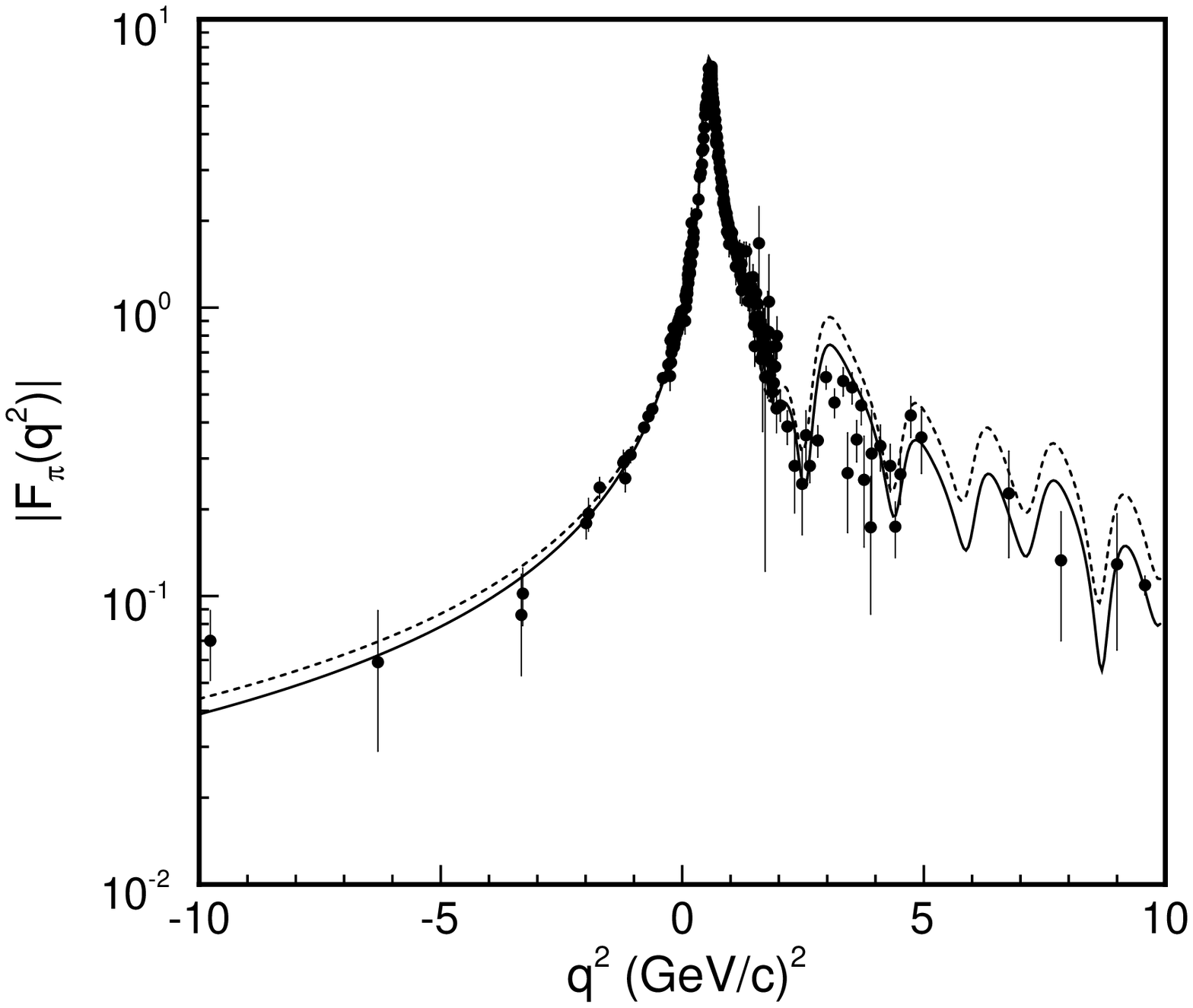}

Figure 3. Pion electromagnetic form factor vs the 
square momentum transfer $q^2$. Dashed and solid lines 
are the results with 
 the asymptotic (see text) and the full pion wave function, respectively. 
Experimental data are from Ref. \cite{baldini}. (After \cite{plbpion}). 
\end{figure} 
Our calculation of the pion form factor contains a very small set of parameters:  
i) the constituent quark mass, ii)  the 
oscillator strength, $\omega$, and iii) the VM 
 widths, $\Gamma_n$, for $M_n >~2.150~GeV$. The up-down quark mass is fixed at 
0.265 $GeV$ \cite{FPZ02}. 
For the first four vector mesons  
the known experimental masses and widths are used in the calculations \cite{pdg}. 
However, the non-trivial $q^2$ dependence of $g^+_{Vn}(q^2)$ in our microscopical model 
implies a shift of the VM masses, with respect to the values  
obtained by using  
Breit-Wigner functions with constant values for $g^+_{Vn}$.  
As a consequence,  
 the value of the $\rho$ meson mass is 
moved  from the usual one, $.775 ~ GeV$, to $.750~GeV$. 
For the  VM, with $M_n >~2.150~GeV$, the mass values corresponding to the model 
of Ref. \cite{FPZ02} 
are used, while for the unknown widths we use a single value 
$\Gamma_n = 0.15 ~GeV$, which presents the best agreement with the compilation of
the experimental data of 
Ref. \cite{baldini}. We consider 20 resonances in our calculations 
to obtain  stability of the results up to $q^2 =~10 ~(GeV/c)^2$.

The oscillator strength is fixed at $\omega = 1.39~GeV^2$  \cite{ani}.  
 The values of the coupling constants, $f_{Vn}$, are evaluated from 
 the model VM  
wave functions through Eq. (\ref{fvn}). The corresponding 
partial decay width, 
$\Gamma_{e^+e-} = 8\pi \alpha^2 ~ f_{Vn}^2 / ( 3 M_n^3 )$   
 (where $\alpha$ is the fine structure constant),  is
 compared for the first three VM with the experimental information in the 
 following table   
\begin{table}[h]
\begin{center} 
\begin{tabular} {|c||c|c|}
\hline~ & ~& ~\\
 VM  & $\Gamma^{th}_{e^+e^-}$  & $\Gamma^{exp}_{e^+e^-}$ \cite{pdg} \\
~ & ~& ~\\
\hline ~ & ~& ~\\
$\rho(770)$  & 6.37 KeV & 6.77 $\pm$ 0.32 KeV \\   
\hline~ & ~& ~\\
$\rho(1450)$  & 1.61 KeV & $>$ 2.30 $\pm$ 0.50 KeV \\  
\hline~ & ~& ~\\
$\rho(1770)$  & 1.23 KeV & $>$ 0.18 $\pm$ 0.10 KeV \\  
\hline
\end{tabular}
\end{center}
\end{table} 

We perform two sets of calculations for the pion form factor. In the first one, we use the 
asymptotic form of the pion valence wave function, obtained  
with $\Lambda_{\pi}(k,P_{\pi}) = 1 $ in Eq. (\ref{wf2});  
 in the second one, we use the eigenstate of 
the square mass operator of Refs. \cite{tobpauli,FPZ02}. The pion radius  
for the 
asymptotic wave function is $r_\pi^{\rm{asymp}}$ = 0.65 fm  
and for the full model wave function is $r_\pi^{\rm{model}}$ = 0.67 
fm, to be compared with the experimental value  
$r_\pi^{\rm{exp}}$ = $0.67 ~\pm ~0.02$ fm \cite{amen}. The good 
agreement with the experimental form factor at low momentum 
transfers is expected, since we have built-in 
the generalized $\rho$-meson dominance. 
 
The calculated pion form factor is shown in Fig. 3 
in a wide region of square 
momentum transfers, from $-10$ $(GeV/c)^2$ up to 10 $(GeV/c)^2$. A general 
qualitative agreement with the data is seen, independently of the detailed form of the pion 
wave function.  
The results obtained with the asymptotic pion wave function and the full model, present 
some differences only above $3 ~(GeV/c)^2$. 
The SL form factor is notably 
well described.
It has to be stressed that the heights of the TL bumps directly depend 
on the calculated values of $f_{Vn}$ and $g^+_{Vn}$. 
   
The introduction of $\omega$-like \cite{Gard} 
and $\phi$-like mesons  
could  improve the description of the data in the TL region.

\section{ THE NUCLEON EM FORM FACTOR IN THE SPACE- AND TIME-LIKE REGIONS: A 
HEURISTIC INTRODUCTION}
A particularly challenging puzzle in the study of hadron em properties 
  is represented by the TL form factors of the nucleon (see e.g.
\cite{Ellis} and \cite{Fenice} for  theoretical and experimental status reports,
respectively). 

For  $q^2=0$ the SU(3) CQ model
yields for the ratio $G_p^M(0)/G_n^M(0)$ the well known result,
i.e. $G_p^M(0)/G_n^M(0)=-3/2$, that can be easily obtained from the expectation
value of a one-body operator, viz. 
\be \hspace{-0.15in}
G_N^M(0) \propto \langle N|~\sigma_z~ \left \{e_u {(1+\tau_3) \over 2} +e_d
 {(1-\tau_3) \over 2}
\right \} |N \rangle=
\nonu = \langle N|~\sigma_z ~\left \{ {1 \over 6} + {\tau_3 \over 2}
\right \} |N \rangle
\label{gone}\ee
where, in the second line, the isoscalar and isovector contributions to the matrix
element are put in evidence. In the calculation of the ratio $G^M_p(0)/G^M_n(0)$ the momentum
distribution of a quark in the nucleon, i.e. the square modulus of the valence
component, is the relevant quantity due to the one-body nature of the
current adopted for
evaluating the ratio in SU(3) CQ model. By using Eq. (\ref{gone}), for the proton we have the contribution
only from a scalar diquark with I=0, while for the neutron we have contributions
from both a (0,0)-diquark  and a (1,1)-diquark, properly weighted for the
spin-isospin part, while the dynamical part turns out to be equal.
\begin{figure}[t]
\includegraphics[width=4.7in]{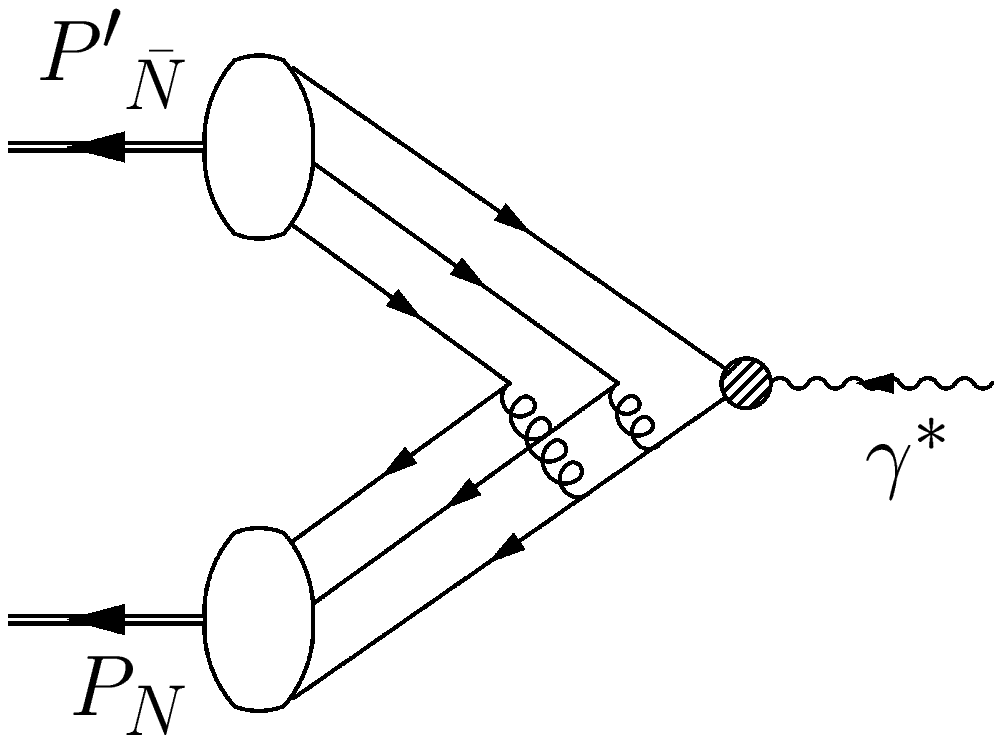}

\vspace{-3.9in}
Figure 4.  Diagrammatic representation of the photon decay 
($\gamma^* \rightarrow N \bar{N}$). \medskip \medskip \medskip 
\end{figure}

  \begin{figure}[b]
\medskip
\includegraphics[width=3.in]{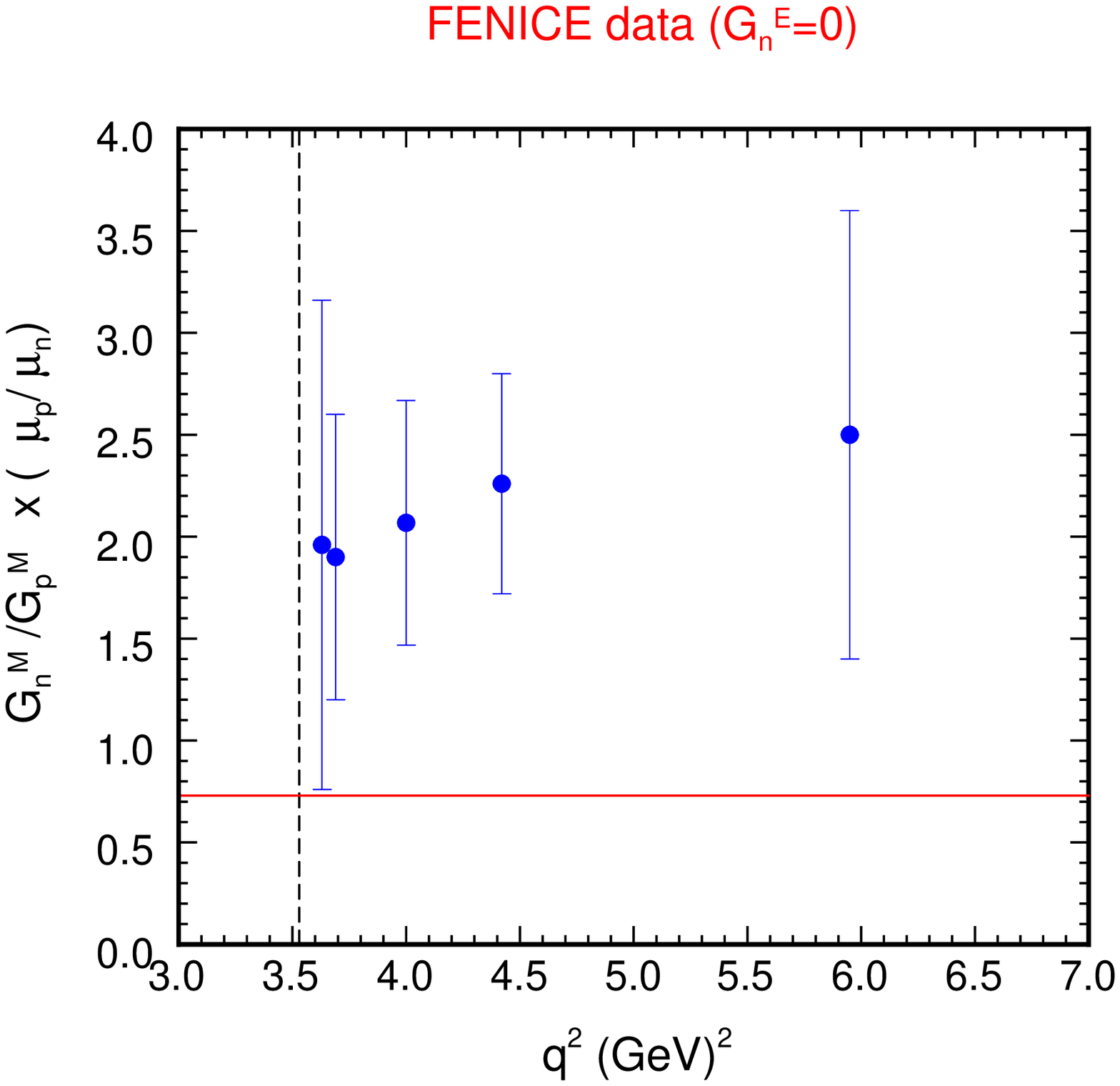}
Figure 5. Solid line: naive expectation  $\equiv~e_d/e_u~\times ~ \mu_p /\mu_n$
(\cite{Ellis}).
Experimental data from ref. \cite{Fenice}. \medskip
\end{figure}

In the TL region, (see Fig. 4), a
  naive perturbative description of the $e^+e^-$ annihilation \cite{Ellis}
  leads to  $|G^M_n(4M^2_n)/G^M_p(4M^2_n)|\equiv~|e_d/e_u| =1/2$, to be
  compared with an experimental value of the order of $1.4$, but with a large error
  bar. As suggested  in \cite{Ellis}, a possible solution is
  given by a strong prevalence of the isovector channel in the $N\bar{N}$
  final state in a two-step production process, where the virtual photon 
  could materialize in a $\rho$-like meson.
  
 In our understanding,   also a dynamical motivation could play a role in
 getting
 close to the experimental result. In the
 TL region, one has to calculate the expectation value 
 of a many-body current,
like in the case of the pion, where the mechanism of the decay of
the virtual photon in a $q\bar{q}$ pair dresses the quark-photon vertex. 
 In order to deal with such a process,  both valence 
 and nonvalence component 
 of the hadron state (see, e.g., Eq. (\ref{ffpi2})) are necessary. This implies
  that the relevant
  quantity becomes the wave function instead of the momentum distribution. 
  It turns out that
a dynamical contribution to the many-body current from 
a (1,1)-diquark   larger than the one from a 
 (0,0)-diquark, produces  an increasing in $G^M_n(4M^2_n)$ with respect to 
 $G^M_p(4M^2_n)$.

 A very preliminary estimate, without taking into account the possible
 difference between the isoscalar and isovector contributions of the 
 quark-photon vertex (as suggested in \cite{Ellis}) and any flavour dependence 
 in the
 diquark-emission vertex,
 indicates a value of  the ratio shown in Fig. 5
 $G^M_n(q^2)/G^M_p(q^2)\times(\mu_p/\mu_n)\sim ~ 1.1$ (while in the
  naive perturbative description one has
  $G^M_n(4M^2_n)/G^M_p(4M^2_n)\equiv~e_d/e_u~\times ~ \mu_p /\mu_n \sim 
0.73$).

\section{SUMMARY}
The theoretical results presented in this contribution show that a VM dominance ansatz 
for the (dressed photon )-($q\overline q$)
vertex, within a CQ model consistent with the meson spectrum, 
is able to give a unified description of the pion form factor both in the SL and
TL regions. 
Using the 
 experimental widths for the first four vector mesons and a single free 
 parameter for the unknown widths of the other vector mesons, the model gives 
 a fair agreement with the TL data, while in the SL region it works 
 surprisingly well. 
 These results encourage the investigation of the TL form factors of the nucleon
 by using  an approach  based on  a simple ansatz for the nonvalence component
 of the nucleon state, following as a  guideline  the pion case.

\end{document}